\documentclass[12pt,a4paper]{article}
\usepackage{fullpage}
\usepackage{amsfonts,amssymb,amsmath,mathrsfs,graphicx,xcolor,bm,mathtools,orcidlink,url,cite}
\definecolor{ultramarine}{rgb}{0.07, 0.04, 0.56}
\definecolor{cadmiumgreen}{rgb}{0.0, 0.42, 0.24}
\definecolor{indigo(dye)}{rgb}{0.0, 0.25, 0.42}
\usepackage{hyperref}
\hypersetup{
colorlinks=true,
citecolor=ultramarine,
linkcolor=cadmiumgreen,
urlcolor=indigo(dye),
}

\begin{document}

\title{Constant-Roll Inflation}
\author{Hayato Motohashi\orcidlink{0000-0002-4330-7024}\\ 
\small Department of Physics, Tokyo Metropolitan University, \\
\small 1-1 Minami-Osawa, Hachioji, Tokyo 192-0397, Japan}

\maketitle

\abstract{Constant-roll inflation is a distinctive class of phenomenological inflationary models in which the inflaton's rate of roll remains constant. It provides an exact solution that is compatible with the latest observational constraints and offers a natural framework for enhancing the curvature power spectrum, which is relevant to the formation of primordial black holes. In this paper, I review constant-roll inflation in memory of Alexei Starobinsky.}

\section{Introduction}
\label{sec:intro}

Alexei Starobinsky's pioneering work in cosmology, astrophysics, and gravitational physics has profoundly shaped our understanding of the universe.
Among his numerous groundbreaking contributions, he played a pivotal role in developing the theory of the early universe.
He proposed one of the first models of cosmic inflation~\cite{BROUT197878,Starobinsky:1980te,Kazanas:1980tx,Sato:1981qmu,PhysRevD.23.347,LINDE1982389,PhysRevLett.48.1220}, which is a cornerstone of modern cosmology, resolving fundamental issues in Big Bang cosmology—such as the horizon, flatness, and monopole problems—through a phase of quasi-exponential expansion of the universe. 
At the same time, it provides a mechanism for generating primordial perturbations~\cite{Starobinsky:1979ty,Mukhanov:1981xt,Starobinsky:1982ee,Bardeen:1983qw,Mukhanov:1985rz,Sasaki:1986hm}, 
which act as the seeds for the large-scale structure of the universe.
He was instrumental in the development of the $\delta N$ formalism~\cite{Lifshitz1960,Starobinsky:1982ee,Salopek:1990jq,Comer:1994np,Sasaki:1995aw,Sasaki:1998ug,Wands:2000dp,Lyth:2003im,Rigopoulos:2003ak,Lyth:2004gb,Lyth:2005fi} and pioneered the stochastic formalism~\cite{Starobinsky:1982ee,Starobinsky:1986fx,Starobinsky:1994bd} to explore generation and evolution of quantum fluctuations during inflation.
He also made key contributions to the theory of reheating~\cite{Kofman:1994rk,Kofman:1997yn,Greene:1997fu}, the process by which the inflaton field oscillates and decays into Standard Model particles after inflation.

I had the privilege of collaborating with Alexei and spending valuable time working with him.  
From 2000 to 2019, he regularly visited the Research Center for the Early Universe (RESCEU) at the University of Tokyo, where I was a PhD student from 2008 to 2013.
Alongside my PhD advisor, Jun'ichi Yokoyama, and Alexei, I explored topics in dark energy~\cite{Motohashi:2009qn,Motohashi:2009zza,Motohashi:2010tb,Motohashi:2010qj,Motohashi:2010sj,Motohashi:2010zz,Motohashi:2011wy,Motohashi:2011odk,Motohashi:2012wh,Motohashi:2012wc} and its connection to inflation~\cite{Motohashi:2012izs}.
Our discussions at RESCEU and at conferences around the world remain unforgettable, deepening both my understanding of fundamental physics and my appreciation for Alexei's unique approach and dedication to research.

Beyond our professional relationship, Alexei was warm and engaging on a personal level as well. 
Outside of physics, we sometimes enjoyed playing table tennis together in the university gym, where I was always impressed by his remarkably steady backhand.

Our collaboration continued after I completed my PhD, shifting focus to inflation~\cite{Motohashi:2014ppa,Motohashi:2017aob,Motohashi:2017vdc,Motohashi:2019tyj}.
We proposed a particular type of inflation, dubbing it `constant-roll inflation', in which the slow-roll condition is replaced by a requirement that the inflaton maintains a constant rate of roll.
This framework generalizes slow-roll inflation, provides an exact solution compatible with the latest observational constraints, and is also relevant to the formation of primordial black holes.
In this paper, dedicated to the memory of Alexei, I shall review constant-roll inflation.

``We should work harder,'' Alexei once said during his banquet address at a conference a decade ago, mourning the untimely passing of his former PhD student Lev Kofman. 
Now, I feel the weight of his words more than ever.
His research legacy continues to inspire us, and each reading of his papers reveals new insights.
It is our responsibility to further advance cosmology and gravitational physics in directions that Alexei could not witness firsthand—but might have foreseen with his remarkable insight.  

\section{Constant-roll inflation}
\label{sec:conroll}

Constant-roll inflation is a phenomenological inflationary model, in which the slow-roll condition is replaced by a requirement that the inflaton's rate of roll remains constant.
It allows for exact analytical solutions and exhibits unique theoretical properties.
Our focus is on constant-roll models in single-field canonical inflation~\cite{Motohashi:2014ppa}, which has been extensively studied~\cite{Martin:2012pe,Motohashi:2017aob,Gao:2017uja,Cicciarella:2017nls,Anguelova:2017djf,Karam:2017rpw,Yi:2017mxs,Gao:2018cpp,Cicoli:2018asa,Morse:2018kda,Cruces:2018cvq,GalvezGhersi:2018haa,Cruces:2018cvq,Gao:2019sbz,Lin:2019fcz,Atal:2019erb,Motohashi:2019rhu,Stein:2022cpk,Davies:2021loj,Karam:2022nym,Motohashi:2023syh,Mishra:2023lhe,Tomberg:2023kli,Tasinato:2023ukp,Wang:2024xdl,Inui:2024sce}.
Extensions beyond single-field or canonical inflation have also been explored~\cite{Hirano:2016gmv,Motohashi:2017vdc,Ito:2017bnn,Mohammadi:2018oku,Gao:2018tdb,Mohammadi:2018wfk,Motohashi:2019tyj,Mohammadi:2020ftb,Guerrero:2020lng,Garnica:2021fuu}.

\subsection{Background exact solution}
\label{sec:back}

We consider a canonical single-field inflation model described by the action
\begin{equation} \label{action-canonical}
S = \int d^4x \sqrt{-g} \left[ \frac{M_{\rm Pl}^2}{2} R - \frac{1}{2} g^{\mu\nu} \partial_\mu\phi\partial_\nu\phi - V(\phi) \right],
\end{equation}
where $M_{\rm Pl}\equiv(8\pi G)^{-1/2}$ is the reduced Planck mass.
We work in a spatially flat Friedmann-Lema\^itre-Robertson-Walker (FLRW) background metric
\begin{align} \label{FLRW}
ds^2=-dt^2+a^2\delta_{ij}dx^idx^j ,
\end{align}
where $a=a(t)$ is the scale factor.
The equations of motion are then given by
\begin{eqnarray} 
3 M_{\rm Pl}^2 H^2 &=& \frac{1}{2} \dot\phi^2 + V, \label{eom1-canonical} \\
-2 M_{\rm Pl}^2 \dot H &=& \dot\phi^2, \label{eom2-canonical} \\
\ddot\phi + 3 H \dot\phi + V' &=& 0 , \label{eom3-canonical}
\end{eqnarray}
where $H\equiv \dot a/a$ is the Hubble parameter, and dots and primes denote derivatives with respect to $t$ and $\phi$, respectively. 
Only two of \eqref{eom1-canonical}--\eqref{eom3-canonical} are independent.

For a given potential $V(\phi)$, one can solve the equations of motion~\eqref{eom1-canonical}--\eqref{eom3-canonical} to obtain the evolution of the scale factor $a(t)$ and the inflaton field $\phi(t)$.
On the other hand, one can require a particular type of inflationary evolution and find a potential that accommodates it.
The constant-roll inflation is a representative example of the latter approach.

Constant-roll inflation is characterized by the following condition~\cite{Martin:2012pe}
\begin{equation} \label{cr-con}
\ddot \phi=\beta H\dot \phi,
\end{equation}
where $\beta$ is a constant. 
This generalizes the slow-roll case, $\ddot \phi/(H\dot \phi) \ll 1$, and the ultra-slow-roll case with $\beta=-3$~\cite{Tsamis:2003px,Kinney:2005vj}. 
While \cite{Martin:2012pe} derived an approximate solution, it was later demonstrated in \cite{Motohashi:2014ppa} that an exact solution satisfying the constant-roll condition can be obtained (see also \cite{Barrow:1990nv,Barrow:1994nt,Contaldi:2003zv,Kofman:2007tr} for earlier related work).

The key to deriving an exact solution 
lies in the Hamilton-Jacobi formalism~\cite{Muslimov:1990be,Salopek:1990jq}, where the Hubble parameter $H$ is treated as a function of $\phi$. 
This approach is valid as long as $t=t(\phi)$ is a single-valued function of $\phi$.
Substituting $\dot H=H'\dot\phi$ into \eqref{eom2-canonical}, we find
\begin{align} \label{HJ}
-2 M_{\rm Pl}^2 H' = \dot\phi.
\end{align}
Taking the time derivative of \eqref{HJ} and using the constant-roll condition \eqref{cr-con}, we obtain
\begin{align} \label{HJ2}
H'' = -\frac{\beta}{2M_{\rm Pl}^2} H.
\end{align}
The general solution $H(\phi)$ to this equation is given by a linear combination of $\exp\left( \pm \sqrt{\frac{-\beta}{2}} \frac{\phi}{M_{\rm Pl}}\right)$.
With this solution $H(\phi)$, we can construct the potential $V(\phi)$ from \eqref{eom1-canonical} and \eqref{HJ} as
\begin{align}
V(\phi) = 3 M_{\rm Pl}^2 H^2 - 2 M_{\rm Pl}^4 H'^2.
\end{align}
Integrating \eqref{HJ}, we obtain the time evolution of $\phi$, allowing us to express $H$ as a function of time and determine the scale factor $a(t)$.

We remark that the above derivation does not rely on the slow-roll approximation.
The slow-roll parameters, defined as
\begin{align} \label{Hsr}
\epsilon_1 \equiv -\frac{d\ln H}{d\ln a}, \qquad \epsilon_{n+1} \equiv \frac{d\ln \epsilon_{n}}{d\ln a},
\end{align}
are not necessarily small in constant-roll inflation.
In fact, in the asymptotic regime $\phi\simeq 0$, corresponding to $t \to \mp \infty$ for $\beta \gtrless 0$, they approach \cite{Motohashi:2014ppa} [see also \eqref{Hsrh} below]
\begin{align} \label{Hsrasymp}
2\epsilon_1 = \epsilon_{2n+1} \simeq 2|\beta|a^{2\beta} \to 0, \qquad \epsilon_{2n} \to 2\beta .
\end{align}
Thus, constant-roll inflation provides a controlled deviation from slow-roll dynamics, offering a versatile framework for understanding inflationary behavior beyond the standard paradigm.
While originally developed as a phenomenological model, the emergence of constant-roll models from string theory has been discussed recently~\cite{Cicoli:2018asa}.

\subsection{Stability and duality}
\label{sec:stab}

Although the Hamilton-Jacobi formalism provides exact solutions for constant-roll inflation, these solutions are not necessarily stable.
Hence, it is crucial to analyze their stability separately. 
The stability analysis reveals intriguing properties and constraints on constant-roll models for different ranges of the parameter $\beta$.

A remarkable feature of constant-roll inflation is the duality between background solutions with $\beta$ and $\alpha \equiv -(3+\beta)$.
As noted in \cite{Motohashi:2014ppa}, the effective inflaton mass squared asymptotically approaches $\alpha\beta M^2$ as $\phi\simeq 0$, and the general solution for the inflaton is the sum of two exponential functions, $e^{\alpha M t}$ and $e^{\beta M t}$, where $M$ is the characteristic energy scale of inflation.
This duality is analogous to the one found in inflation with a quadratic potential~\cite{Tzirakis:2007bf}. 
In fact, the constant-roll potential can be approximated by a quadratic form near its extremum, leading to the same duality.

It is important to note that the construction of the constant-roll solution provides only the latter solution, $e^{\beta M t}$, which is consistent with the condition~\eqref{cr-con}.
Although the Hamilton-Jacobi formalism produces only one solution, each constant-roll solution with a given $\beta$ has a dual counterpart with $\alpha$. 
The stability properties of these dual solutions can differ. 
The attractor behavior of the inflationary dynamics is governed by one of the two solutions, while the other serves as a non-attractor solution, making the constant-roll condition unstable for that branch.

It has been explicitly demonstrated that constant-roll inflation is unstable for certain parameter regions~\cite{Morse:2018kda,Gao:2019sbz,Lin:2019fcz}. 
For $\beta < -3/2$, numerical studies show that $\ddot\phi/(H\dot\phi)$ asymptotically approaches $-(3+\beta)$, indicating that the constant-roll analytical solution is not an attractor, including the ultra-slow-roll case with $\beta = -3$. 
Conversely, for $-3/2 < \beta$, the constant-roll solution is stable and acts as the attractor.

The duality has been further investigated and provided rich phenomenology in the analysis of primordial perturbations~\cite{Wands:1998yp,Leach:2000yw,Starobinsky:2005ab,Tzirakis:2007bf}. 
The duality in the quadratic potential was recently revisited in \cite{Pi:2022ysn} as a logarithmic duality of curvature perturbations.
Extensions of this concept to constant-roll inflation have clarified characteristic non-Gaussian tails~\cite{Wang:2024xdl,Inui:2024sce}.
The implications of logarithmic non-Gaussianity for primordial black holes and induced gravitational waves have also been studied recently~\cite{Inui:2024fgk,Shimada:2024eec}.

\subsection{Perturbations}
\label{sec:pert}

Let us consider the gauge invariant curvature perturbation, $\zeta$, related to the metric perturbation through $\delta g_{ij} = a^2(1-2\zeta)\delta_{ij}$ in a gauge $\delta\phi=0$. 
The evolution of the mode function $v_k\equiv \sqrt{2}M_{\rm Pl} z \zeta_k$ with 
$z\equiv a \sqrt{\epsilon_1}$ is governed by the Mukhanov-Sasaki equation~\cite{Mukhanov:1985rz,Sasaki:1986hm}:
\begin{align} \label{mseq} \frac{d^2v_k}{d\tau^2}+\left(k^2-\frac{1}{z}\frac{d^2z}{d\tau^2} \right) v_k=0, \end{align}
where $\tau$ is the conformal time defined by $d\tau =dt/a$.
The potential term $\frac{1}{z}\frac{d^2z}{d\tau^2}$ is exactly expressed in terms of slow-roll parameters:
\begin{align} \frac{1}{z}\frac{d^2z}{d\tau^2}=a^2H^2\left(2-\epsilon_1+\frac{3}{2}\epsilon_2+\frac{1}{4}\epsilon_2^2-\frac{1}{2}\epsilon_1\epsilon_2+\frac{1}{2}\epsilon_2\epsilon_3 \right). \end{align} 

The asymptotic expression of the slow-roll parameters~\eqref{Hsrasymp} allows us to derive a simple relation
\begin{align} \frac{1}{z}\frac{d^2z}{d\tau^2} \simeq \frac{\nu^2-1/4}{\tau^2}, \qquad \nu \equiv |\beta+3/2|. \end{align} 
and then the spectral index $n_s$ of the curvature power spectrum evaluated at horizon exit is given by~\cite{Motohashi:2014ppa}:
\begin{align} \label{nsm1}
n_s-1 = 3-|2\beta+3|.
\end{align}
Thus, the spectrum generated by constant-roll inflation is 
scale-invariant for $\beta=0$ or $-3$,
red-tilted for $\beta < -3$ or $\beta > 0$, 
and blue-tilted for $-3<\beta<0$.
We can also see that the same spectral index is predicted by background solutions with $\beta$ and $\alpha \equiv -(3+\beta)$.
Below, we shall explore the properties of red-tilted and blue-tilted models in greater detail.

An important aspect of constant-roll inflation is the behavior of the curvature perturbation $\zeta$ on superhorizon scales. 
The general solution of the Mukhanov-Sasaki equation~\eqref{mseq} in this regime is given by
\begin{align} \label{superH}
\zeta = A + B \int \frac{dt}{a^3\epsilon_1},
\end{align}
where $A$ and $B$ are constants. 
The first term remains constant, whereas the second term evolves in time. 
In standard slow-roll inflation, where $\epsilon_1$ is nearly constant and small, this second term decays, causing the curvature perturbation to freeze on superhorizon scales. 
However, in more general case such as constant-roll inflation, the evolution of $\epsilon_1$ can alter this behavior. 
Since the constant-roll condition~\eqref{cr-con} implies $\dot\phi \propto a^\beta$, it follows that $\epsilon_1 = \dot\phi^2/(2H^2) \simeq a^{2\beta}$.
For $\beta < -3/2$, the second term in \eqref{superH} grows, leading to a superhorizon growth of curvature perturbations. 
The ultra-slow-roll case ($\beta=-3$) is a well-known example of this phenomenon.

\begin{figure}[t]
\centering
\includegraphics[width=.99\textwidth]{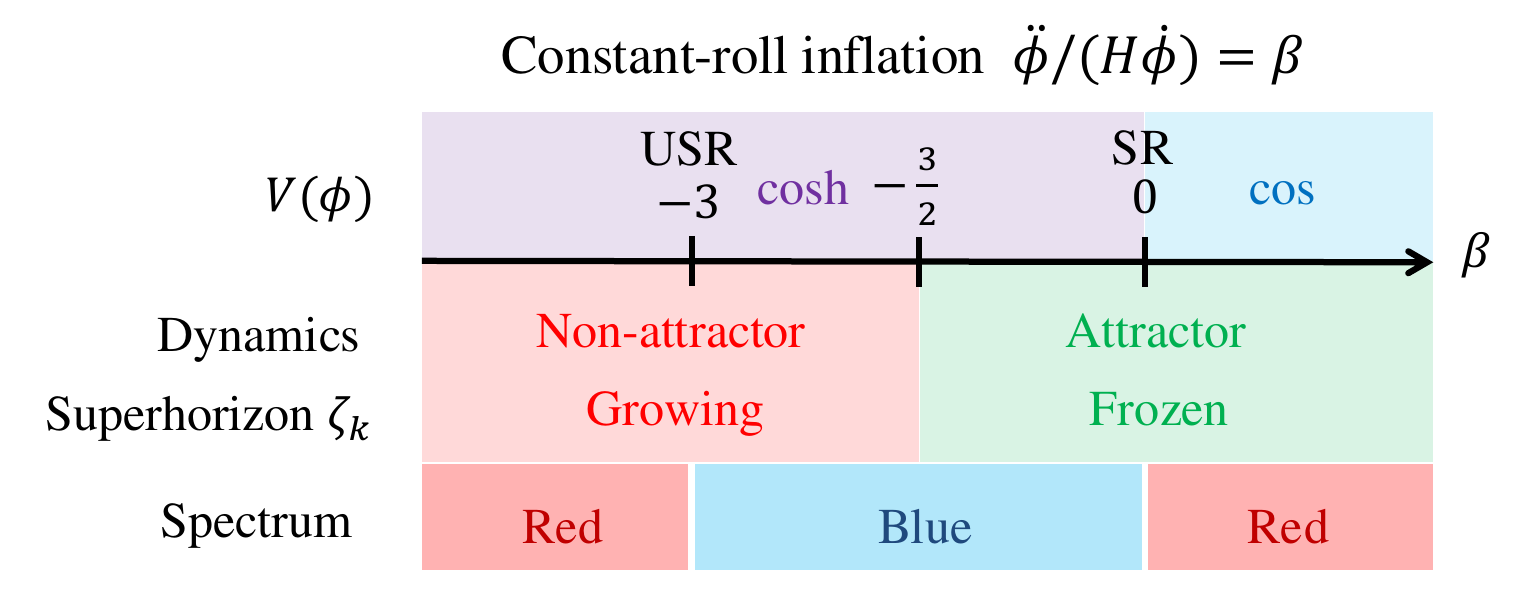}
\caption{Classification and properties of constant-roll models.}
\label{fig:beta}
\end{figure}

Constant-roll models can be classified based on the value of $\beta$, as illustrated in Fig.~\ref{fig:beta}. 
Each regime exhibits distinct characteristics in terms of the potential shape, stability properties, superhorizon perturbation evolution, and spectral tilt. 

In the following sections, we focus on attractor constant-roll models with $\beta > -3/2$. 
We examine the implications of a red-tilted spectrum for $\beta > 0$ and a blue-tilted spectrum for $-3/2 < \beta < 0$.

\subsection{Red-tilted spectrum}
\label{sec:red}

Let us first consider the case of positive $\beta$.
There are several potentials that maintain the constant-roll condition.
Among them, a viable constant-roll potential that generates a red-tilted spectrum consistent with observational constraints is given by~\cite{Motohashi:2014ppa}
\begin{align} \label{crpot-red} V(\phi)= 3M^2M_{\rm Pl}^2 \left[ 1-\frac{3+\beta}{6}\left\{1-\cos \left(\sqrt{2\beta } \frac{\phi}{M_{\rm Pl}} \right) \right\} \right] , \end{align} 
where $M$ is the energy scale of inflation. 
The exact solutions for $a(t)$, $H(t)$ and $\phi(t)$ are given by
\begin{align}
\phi&=  2\sqrt{\frac{2}{\beta }}M_{\rm Pl} {\rm arctan} (e^{\beta Mt}), \\
H&= -M\tanh (\beta Mt) = M\cos\left( \sqrt{\frac{\beta }{2}} \frac{\phi}{M_{\rm Pl}} \right) , \\
a&\propto \cosh^{-1/\beta } (\beta Mt) = \sin^{1/\beta } \left( \sqrt{\frac{\beta }{2}} \frac{\phi}{M_{\rm Pl}} \right) .
\end{align}

\begin{figure}[t]
\centering
\includegraphics[width=.7\textwidth]{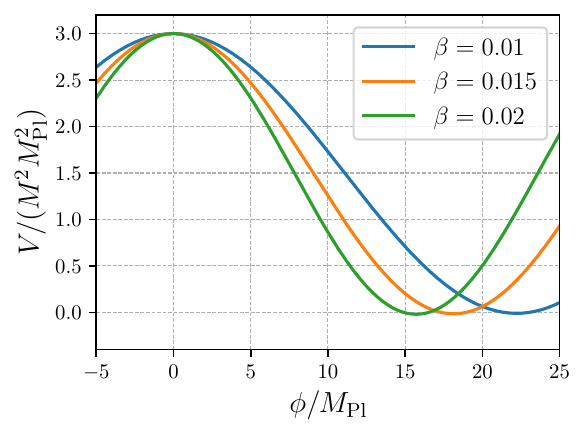}
\caption{Constant-roll potential~\eqref{crpot-red} for $\beta>0$.}
\label{fig:CRpot_positive}
\end{figure}

The potential is shown in Fig.~\ref{fig:CRpot_positive} for representative parameters.
In the following arguments we focus on the region $\phi\geq 0$.
The inflaton asymptotically approaches the origin in the infinite past, and rolls down the potential in the positive direction as time progresses.

This potential resembles natural inflation~\cite{Freese:1990rb} but includes an additional negative cosmological constant, $-M^2\beta$. 
The critical field value $\phi_c$ where $V=0$ is given by
\begin{align}  \phi_c = \frac{M_{\rm Pl}}{\sqrt{2\beta }} \arccos \left( 1-\frac{6}{3+\beta} \right)
= M_{\rm Pl}\sqrt{\frac{2}{\beta} } \arcsin\sqrt{\frac{3}{3 + \beta} } .
\end{align}
To ensure a graceful exit from constant-roll inflation, the potential must be truncated before reaching $V<0$ at a cutoff field value $\phi_0$, which satisfies $\phi_0<\phi_c$. 
After $\phi_0$ we assume a certain kind of phase transition that ends inflation.
The cutoff can be chosen such that $\phi_0 \ll \phi_c$ or $\phi_0 \lesssim \phi_c$. 

Let us denote $\phi_i$ as a field position where the generation of perturbations relevant to CMB scales occurs. 
This should occur sufficiently early, say $60$ e-folds before inflation ends at $\phi_0$.
Consequently, regardless of whether the cutoff $\phi_0$ is far from or close to the critical value $\phi_c$, CMB scale perturbations are expected to be generated at $\phi=\phi_i \ll \phi_c$.
In this region, the potential closely resembles that of quadratic hilltop inflation~\cite{Boubekeur:2005zm}.

The number of e-folds counted from $\phi=\phi_c$ is given by
\begin{align} \label{efold} N \equiv \ln\left[\frac{a(\phi)}{a(\phi_c)}\right] = \frac{1}{\beta} \ln \left[\sqrt{\frac{3+\beta}{3}} \sin \left( \sqrt{\frac{\beta }{2}} \frac{\phi}{M_{\rm Pl}} \right) \right] .\end{align}
Note that $N$ is negative and increasing for $\phi<\phi_c$, meaning it does not measure the total duration of inflation but rather the e-folds until the critical field value $\phi_c$ where $V=0$. 

The CMB scale perturbations are generated at $\phi=\phi_i$ and inflation ends at some earlier point $\phi=\phi_0$, both of which are smaller than $\phi_c$. 
While $\phi_c$ is fixed for a given $\beta$, $\phi_0$ is not fixed and can be chosen any value smaller than $\phi_c$ in general.
If one assumes that the CMB pivot scale exited the horizon $60$ e-folds before the end of inflation, then $N$ corresponding to the CMB scales satisfies $N\leq -60$ but not necessarily $N=-60$.
Therefore, in addition to $M$ and $\beta$, we have an additional parameter $\phi_i$ (or equivalently $N$) in the constant-roll model, and the prediction of CMB observables depends on these three parameters.

In \cite{Motohashi:2017aob}, the slow-roll approximation was used to evaluate inflationary observables:
\begin{align}
n_s-1 &\approx -6\epsilon_V+2\eta_V, \\
r&\approx 16\epsilon_V,\\
\frac{dn_s}{d\ln k}&\approx 16\epsilon_V \eta_V-24\epsilon_V^2-2\xi_V.
\end{align}
where the potential slow-roll parameters are defined as
\begin{align}
\epsilon_V \equiv \frac{1}{2} \left(\frac{V'}{V}\right)^2, \quad 
\eta_V\equiv \frac{V''}{V}, \quad 
\xi_V\equiv \frac{V'V''}{V^2} ,
\end{align}
and to constrain model parameters $\beta$ and $\phi_i$.
It was found that $\beta\approx 0.015$ produces a red-tilted spectrum consistent with CMB constraints~\cite{Motohashi:2014ppa,Motohashi:2017aob}.

We can also consider an approximation based on the Hubble-flow slow-roll parameters defined by \eqref{Hsr}.
Since the Hubble parameter is expressed as 
\begin{align} H = M \sqrt{ 1 - h   } , \qquad h \equiv \frac{3 e^{2 \beta N}}{3 + \beta }, \end{align}
the Hubble-flow slow-roll parameters~\eqref{Hsr} are expressed as
\begin{align}
2\epsilon_1=\epsilon_{2n+1} = \frac{2\beta h}{1-h}, \qquad 
\epsilon_{2n} = \frac{2\beta}{1-h} . \label{Hsrh}
\end{align}
Applying the Hubble-flow slow-roll approximation yields
\begin{align}
\Delta_\zeta^2 &\approx \frac{H^2}{8\pi^2 \epsilon_1} = \frac{M^2}{8\pi^2}\frac{(1-h)^2}{\beta h} , \label{hsr1} \\
n_s-1 &\approx -2 \epsilon_1 - \epsilon_2 = - 2\beta \frac{1+h}{1-h}, \label{hsr2} \\
r&\approx 16\epsilon_1 = 16\beta \frac{ h}{ 1 -h  } . \label{hsr3}
\end{align}
In general, the Hubble-flow slow-roll approximation works better than the potential slow-roll approximation as the latter assumes the approximate relations between the slow-roll parameters such as $\epsilon_1\approx \epsilon_V$.
The formulae~\eqref{hsr1}--\eqref{hsr3} directly relate the observables to 
$M$, $\beta$, and $N$, the number of e-folds~\eqref{efold} counted from the critical field value $\phi=\phi_c$.

We see that there is a degeneracy between $M$ and $N$ since $M$ only appears through the combination $M_s \equiv e^{-\beta N} M$ in \eqref{hsr1}--\eqref{hsr3}.
Therefore, the amplitude of the curvature power spectrum constrains the combination $M_s$ rather than $M$~\cite{GalvezGhersi:2018haa}.

As noted in \cite{Motohashi:2014ppa}, constant-roll inflation allows for a much smaller tensor-to-scalar ratio $r$ than other scenarios due to the freedom in choosing $N$ (or equivalently $\phi_i$). 
By taking $|N|$ sufficiently large (or $\phi_i$ sufficiently small), $r$ can be arbitrarily small.
This topic was recently highlighted in \cite{Stein:2022cpk}, demonstrating that there is no lower bound on $r$ for simple single-field inflation models.

\begin{figure}[t]
\centering
\includegraphics[width=.7\textwidth]{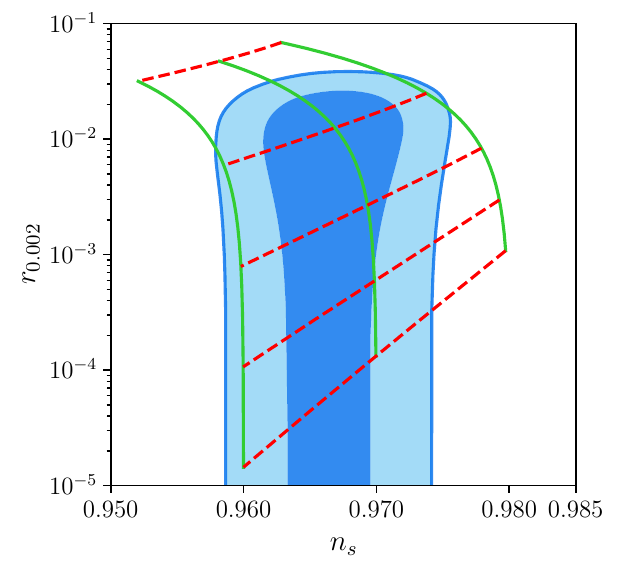}
\caption{Spectral index and tensor-to-scalar ratio predicted from constant-roll inflation assuming the Hubble-flow slow-roll approximation.  
$\beta= 0.010, 0.015, 0.020$ (green solid, from right to left) and $N=-60, -100, -150, -200, -250$ (red dashed, from top to bottom) are shown.
Blue contours indicate the latest CMB constraints from Planck 2018, BICEP/Keck 2018, and BAO~\cite{Planck:2018jri,BICEP:2021xfz}, reproduced using the publicly available data and Python code released by the BICEP/Keck Collaboration (\url{http://bicepkeck.org/bk18_2021_release.html}).}
\label{fig:constraint}
\end{figure}

Assuming the Hubble-flow slow-roll approximation, in Fig.~\ref{fig:constraint}, we compare the predicted spectral index and tensor-to-scalar ratio against constraints from Planck 2018, BICEP/Keck 2018, and BAO~\cite{Planck:2018jri,BICEP:2021xfz}.
The constant-roll inflation with $\beta\approx 0.015$ is consistent with the observational constraints.
It covers the parameter region in a unique way and accommodates small $r$.
While this result is based on the slow-roll approximation and in general deviates from numerical evaluations~\cite{Gao:2018cpp,GalvezGhersi:2018haa}, the qualitative behavior is robust.

The constant-roll inflation with a red-tilted spectrum serves as a simple, exactly solvable, and observationally viable model.
It provides a unique prediction of the spectral index $n_s$ and tensor-to-scalar ratio $r$, especially accommodating negligible $r$.
Future observations will further constrain $r$, allowing this scenario to be tested in the near future.

\subsection{Blue-tilted spectrum}
\label{sec:blue}

\begin{figure}[t]
\centering
\includegraphics[width=.7\textwidth]{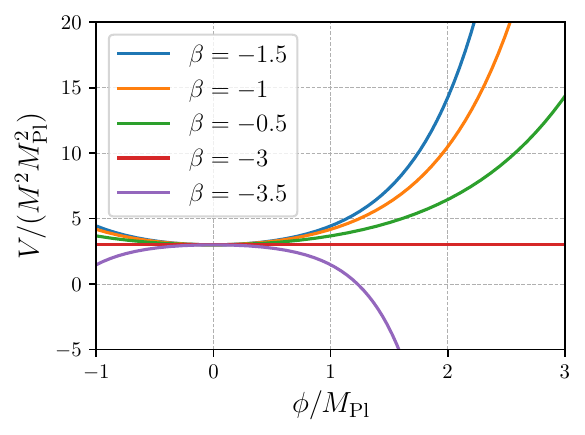}
\caption{Constant-roll potential~\eqref{crpot-blue} for $\beta<0$.}
\label{fig:CRpot_negative}
\end{figure}

Next, let us consider the case of negative $\beta$.
In this case, the constant-roll potential is given by~\cite{Motohashi:2014ppa}
\begin{equation} \label{crpot-blue}
V(\phi) = 3 M^2M_{\rm Pl}^2\left[1-\frac{3 + \beta}{6}\left\{1- \cosh\left( \sqrt{2|\beta|}\frac{\phi}{M_{\rm Pl}}\right)\right\}\right],\end{equation}
and the corresponding exact solution for the background evolution is given by 
\begin{align} 
\phi &= M_{\rm Pl} \sqrt{\frac{2}{|\beta|}}\ln\left[\coth\left(\frac{|\beta|}{2}M t\right)\right],\\
H &= M \coth (|\beta| Mt) = M \cosh \left( \sqrt{\frac{|\beta|}{2}} \frac{\phi}{M_{\rm Pl}} \right),\\
a&\propto \sinh^{1/|\beta| } (|\beta| Mt) = \sinh^{-1/|\beta| } \left( \sqrt{\frac{|\beta|}{2}} \frac{\phi}{M_{\rm Pl}} \right) .
\end{align}

We show the potential in Fig.~\ref{fig:CRpot_negative} for representative parameters.
In the following arguments we focus on the region $\phi\geq 0$.
For $\beta=-3$, the potential becomes constant, reducing to ultra-slow-roll inflation. 
The potential is convex for $-3<\beta<0$ and concave for $\beta<-3$.
For the following, we focus on the regime $-3/2<\beta<0$, for which the above background solution is the attractor, and superhorizon curvature perturbations freeze.
The inflaton monotonically rolls down the potential, asymptotically approaching the origin in the infinite future.

As seen from \eqref{nsm1}, this model generates a blue-tilted spectrum, making it particularly relevant to the formation of primordial black holes (PBHs)~\cite{Zeldovich:1967lct,Hawking:1971ei,Carr:1974nx,Carr:1975qj}, which may form from the direct collapse of overdense regions in the early universe due to large primordial perturbations on small scales
(for recent reviews, see \cite{Sasaki:2018dmp,Carr:2020gox,Escriva:2022duf}).

\begin{figure}[t]
\centering
\includegraphics[width=.7\textwidth]{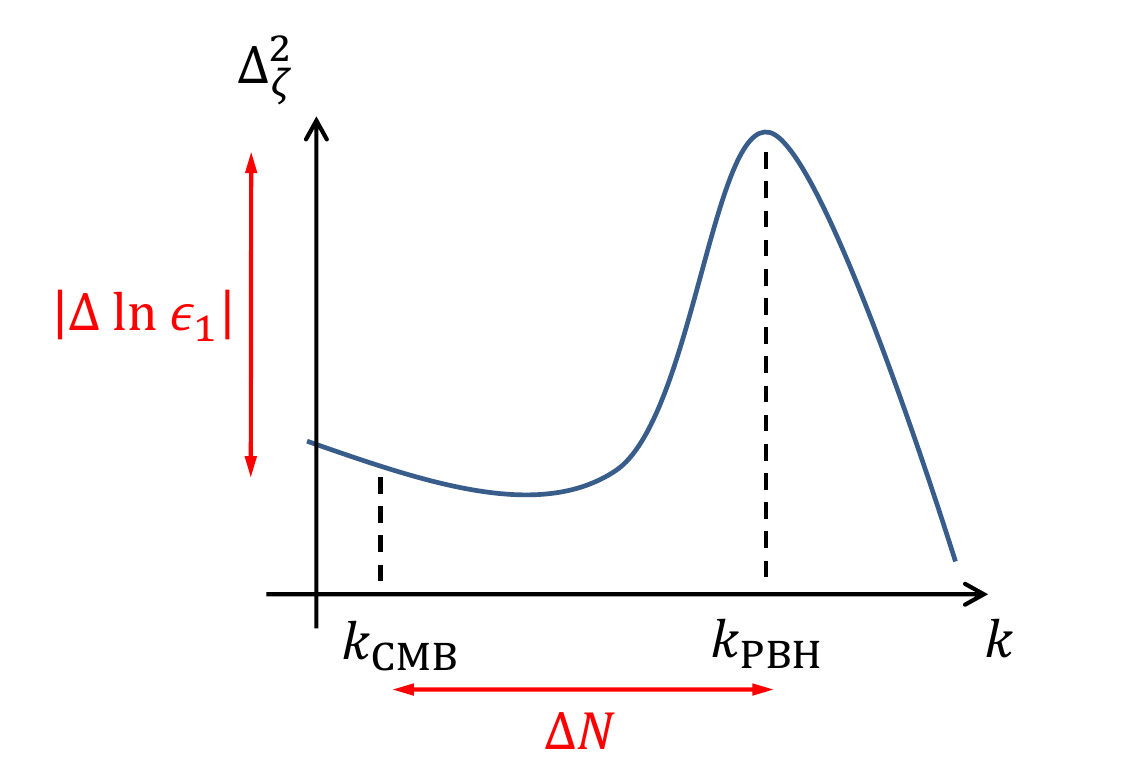}
\caption{Schematic representation of the no-go theorem of slow roll in PBH production.}
\label{fig:nogoSR}
\end{figure}

Constant-roll inflation provides a natural framework for PBH-generating models in connection with the no-go theorem of slow roll for PBH production in canonical single-field inflation~\cite{Motohashi:2017kbs}.
This theorem states that to generate a peak in the curvature power spectrum relevant to PBH formation, a transient violation of the slow-roll condition is necessary. 
This is because the amplitude of the curvature power spectrum, constrained to $\mathcal{O}(10^{-9})$ on CMB scales, requires a significant enhancement to $\mathcal{O}(10^{-2})$ within $60$ efolds, as depicted in Fig.~\ref{fig:nogoSR}.

The slow-roll violation necessary for PBH production is quantified by~\cite{Motohashi:2017kbs}
\begin{align} \label{nogo1}
\frac{\Delta \ln \epsilon_1}{\Delta N} \lesssim -0.4 . 
\end{align}
where $\Delta N$ corresponds to the number of efolds between CMB and PBH scales, and $\Delta \ln \epsilon_1$ is the corresponding change in the slow-roll parameter.
This bound corresponds to the scenario where the smallest mass (or equivalently the largest $\Delta N$) PBHs constitute all the dark matter.

Stronger slow-roll violations are required for different PBH scenarios. 
For instance, if PBHs are responsible for the observed microlensing events towards the Galactic bulge, 
PBHs are roughly in the Earth-Jupiter mass range, $\mathcal{O}(10^{28})\ {\rm g}$, and constitute $\mathcal{O}(1)$\% of dark matter~\cite{Niikura:2019kqi}.
The bound in this case tightens to $\Delta \ln \epsilon_1/\Delta N \lesssim -0.7$.
On the other hand, if PBHs are responsible for gravitational wave events 
detected by LIGO-Virgo-KAGRA collaborations~\cite{Abbott:2016blz,LIGOScientific:2014pky,VIRGO:2014yos,KAGRA:2020tym}, the mass is $\mathcal{O}(10) M_\odot$ and the abundance is $\mathcal{O}(0.1)$\% of dark matter~\cite{Sasaki:2016jop}.
The bound further strengthens to $\Delta \ln \epsilon_1/\Delta N \lesssim -1$.

Although the above estimations are based on various approximations carrying large uncertainties, they all enter into $\Delta \ln \epsilon_1/\Delta N$ logarithmically. 
Even orders of magnitude changes in the mass scale and power spectrum amplification would not qualitatively change this result.

These constraints inform viable inflationary models for PBH formation.
The bound on $\Delta \ln \epsilon_1/\Delta N$ implies that the amplitude of the second slow-roll parameter, $\epsilon_2 = d \ln \epsilon_1/d N$, should be large.
The simplest scenario would be to consider a constant but nonnegligible value of $\epsilon_2$, which is where constant-roll models come into play.
Such a constant-roll phase responsible for the significant enhancement of curvature power spectrum should be sandwiched between different phases of inflation, which are responsible for generating CMB scale perturbations and preventing overproduction of PBHs.
Thus, we are interested in a transient phase of blue-tilted constant-roll inflation.

A special case $\epsilon_2=-6$ corresponds to the ultra-slow-roll inflation with the flat plateau potential.
The PBH production scenarios from transient ultra-slow-roll inflation or inflection models have been actively investigated~\cite{Garcia-Bellido:2017mdw,Ezquiaga:2017fvi,Germani:2017bcs,Motohashi:2017kbs,Ballesteros:2017fsr,Cai:2018dkf,Cicoli:2018asa,Biagetti:2018pjj,Dalianis:2018frf,Pattison:2018bct,Rasanen:2018fom,Byrnes:2018txb,Passaglia:2018ixg,Bhaumik:2019tvl,Cheong:2019vzl,Ragavendra:2020sop,Tasinato:2020vdk,Figueroa:2020jkf,Pattison:2021oen,Inomata:2021uqj,Inomata:2021tpx,Geller:2022nkr,Karam:2022nym,Caravano:2024moy,Fujita:2025imc}. 
However, ultra-slow-roll inflation, despite its simplest constant potential, is not straightforward. 
In terms of slow-roll violation, ultra-slow-roll inflation represents an extreme case, with $\epsilon_2 = -6$, far exceeding the minimum violation~\eqref{nogo1}, $\epsilon_2 \lesssim -0.4$, required for PBH production.
It features a non-attractor phase where superhorizon perturbations grow, leading to diffusion effects. 
These features require special care in analysis.

\begin{figure}[t]
\centering
\includegraphics[width=.49\textwidth]{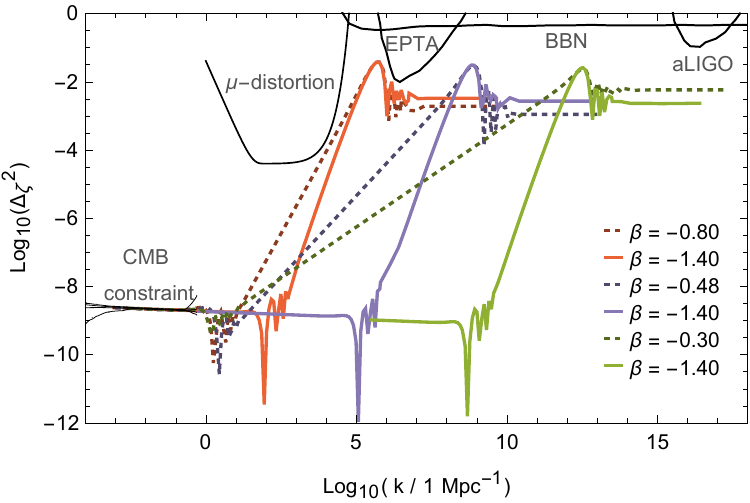}
\includegraphics[width=.49\textwidth]{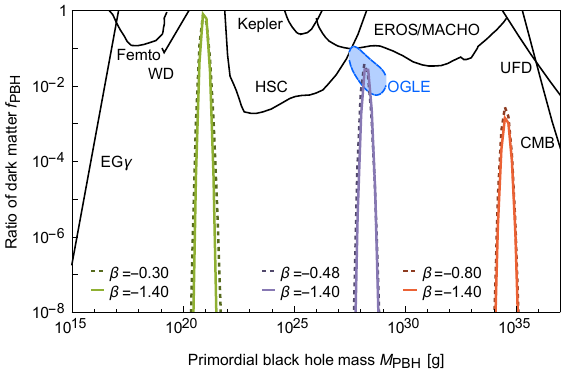}
\caption{Enhancement of the curvature power spectrum (left) and the resulting PBH abundance (right) from transient constant-roll inflation. 
Reprinted with permission from \cite{Motohashi:2019rhu}. \copyright 2020 IOP Publishing Ltd and Sissa Medialab. All rights reserved.}
\label{fig:PBH}
\end{figure}

A more stable alternative is transient constant-roll inflation~\cite{Cicoli:2018asa,Cruces:2018cvq,Atal:2019erb,Motohashi:2019rhu,Mishra:2019pzq,Atal:2019erb,Ozsoy:2020kat,Davies:2021loj,Karam:2022nym,Motohashi:2023syh,Mishra:2023lhe}
which naturally accommodates slow-roll violations necessary for PBH formation while maintaining an attractor behavior for $-3/2<\beta<0$.
A simple piecewise model that realizes a transient constant-roll phase between two slow-roll phases is demonstrated in \cite{Motohashi:2019rhu},
where the spectral tilt of the enhancement can be controlled by the constant-roll parameter, providing greater flexibility and various PBH production scenarios, as illustrated in Fig.~\ref{fig:PBH}.
The non-Gaussianity and one-loop corrections in such models have also been explored recently~\cite{Motohashi:2023syh,Inui:2024sce}.
Beyond exact constant-roll solution, approximate constant-roll behavior can be realized in potentials with bump-like features.
In such models, a transient nearly-constant-roll phase leads to a localized enhancement of the power spectrum~\cite{Mishra:2019pzq,Atal:2019erb}. 

A transient phase of blue-tilted constant-roll inflation provides a more flexible alternative to transient ultra-slow-roll scenarios.
Its advantages make the model a compelling framework for PBH formation, providing both theoretical robustness and observational viability. 
Future studies, including refined non-Gaussianity analyses and constraints from gravitational wave experiments, will further test the predictions of this model.

\section{Conclusion}
\label{sec:con}

Similar to many other models proposed by Starobinsky, constant-roll inflation provides a simple yet elegant framework with valuable insights, offering a rich playground for exploring inflationary dynamics beyond the standard slow-roll paradigm.
It allows for an exact solution where the inflaton maintains a constant rate of roll.
Depending on the parameter branches, constant-roll inflation can generate either a red-tilted spectrum of curvature perturbations, consistent with the latest observational constraints, or a blue-tilted spectrum conducive to the generation of PBHs.  
In both cases, future observations and theoretical developments will be essential for testing the model and deepening our understanding of the early universe.

\section*{Acknowledgements}
I am deeply grateful to Alexei A.\ Starobinsky for his collaborations, insightful discussions, and invaluable support over the years, as well as his friendship I will always cherish.
I would also like to thank my collaborators,  
Wayne Hu,
Ryoto Inui,
Cristian Joana,
Jerome Martin,
Shinji Mukohyama, 
Michele Oliosi,
Samuel Passaglia,
Shi Pi,
Teruaki Suyama,
Yuichiro Tada,
Jun'ichi Yokoyama, and 
Shuichiro Yokoyama
for the fruitful collaborations on the topics discussed in this paper.  
This work was supported by 
JSPS KAKENHI Grant No.~JP22K03639.

\appendix

\section*{Appendix}
\label{sec:dS}

In the main text, we discuss that the key to constructing exact solutions in constant-roll inflation is the Hamilton-Jacobi formalism.
This approach is also useful for obtaining exact de Sitter constant-roll solutions in scalar-tensor gravity, as demonstrated in \cite{Motohashi:2019tyj}.

In this Appendix, we present additional examples of exact de Sitter solutions with a non-constant scalar field, including those beyond constant-roll models. 
These cases were originally discussed and considered for inclusion in \cite{Motohashi:2019tyj} but were ultimately left out.

\subsection*{Exact de Sitter solutions}

Let us consider the following scalar-tensor theory with a non-minimal coupling of a scalar field to gravity:
\begin{equation} \label{action}
S = \int d^4x \sqrt{-g} \left[ \frac{1}{2} F(\phi)R - \frac{1}{2} g^{\mu\nu} \partial_\mu\phi\partial_\nu\phi - U(\phi) \right].
\end{equation}
With the spatially flat FLRW background metric~\eqref{FLRW}, the equations of motion for the action~\eqref{action} are given by
\begin{eqnarray} 
&& 3 F H^2 = \frac{1}{2} \dot\phi^2 + U - 3 H \dot F , \label{eom1} \\
&& -2 F \dot H = \dot\phi^2 + \ddot F - H \dot F, \label{eom2} \\
&& \ddot\phi + 3 H \dot\phi + U' - 3 (\dot H + 2 H^2) F' = 0 . \label{eom3}
\end{eqnarray}

We can derive the exact de Sitter solution with a constant scalar field as follows.
Substituting constant profile $H=H_0$ and $\phi=\phi_0$ to \eqref{eom1} and \eqref{eom3}, we obtain 
\begin{eqnarray}
3 F(\phi_0) H_0^2 &=& U(\phi_0), \label{phi0-1} \\
6 F'(\phi_0) H_0^2 &=& U'(\phi_0). \label{phi0-2}
\end{eqnarray}
Namely, $\phi_0$ for the de Sitter solution is obtained as a root of the following algebraic equation:
\begin{equation}
\left. \left( \frac{U}{F^2} \right)' \right|_{\phi=\phi_0} = 0,
\end{equation}
and then $H_0$ is determined by \eqref{phi0-1} or \eqref{phi0-2}.

Let us derive all the exact de Sitter solutions with a non-constant scalar field assuming $H=H_0$ and $\dot\phi\neq 0$, which is a more nontrivial task.
We adopt a Hamilton-Jacobi formalism and regard $\dot\phi$ as a function of $\phi$:
\begin{equation}
\dot\phi=A(\phi).
\end{equation}
Then, \eqref{eom1} and \eqref{eom2} are rewritten as
\begin{eqnarray}
&A'+\dfrac{F''+1}{F'}A = H_0, \label{Aeq} \\
&U = 3 H_0^2 F + 3 H_0 F'A  - \frac{1}{2} A^2 . \label{Ueq}
\end{eqnarray}
Using the method of integrating factors, we can solve the differential equation \eqref{Aeq} for $A(\phi)$ to obtain 
\begin{equation} \label{Asol}
A(\phi) = \frac{H_0}{F'}\exp\left(-\int\frac{d\phi}{F'}\right) \int d\phi F' \exp\left(\int\frac{d\phi}{F'}\right).
\end{equation}
Plugging this into \eqref{Ueq}, we obtain the potential $U(\phi)$ that accommodates the exact de Sitter solutions.
Note that this branch shows up only when $F' \ne 0$, i.e., when the scalar field non-minimally couples to gravity. 
The de Sitter branch with $\dot\phi\ne 0$ is not possible in the case of Einstein-Hilbert action with a canonical scalar field.

The de Sitter solution~\eqref{Asol} includes both constant-roll and non-constant-roll solutions since it does not satisfy the constant-roll condition~\eqref{cr-con} in general. 
Indeed, we see from \eqref{Asol} that 
\begin{equation} \label{ratio} 
\frac{\ddot\phi}{H_0\dot\phi}=\frac{A'}{H_0} 
= 1-\frac{F''+1}{F'^2} \exp\left(-\int\frac{d\phi}{F'}\right) \int d\phi F' \exp\left(\int\frac{d\phi}{F'}\right),
\end{equation}
which is not constant in general.
From \eqref{ratio}, we can derive a simple criterion:
The exact de Sitter solution \eqref{Asol} is also constant-roll one when $\delta$ is constant, which we define as
\begin{equation} \label{delta}
\delta = \frac{(F''+1)^2}{-F'''F'+F''(2F''+3)+1},
\end{equation}
and $\delta=1-\beta$ holds when it is constant.

\subsection*{Non-minimally coupled scalar field}

Now, let us focus on a particular subclass, the case of non-minimally coupled scalar field with 
\begin{equation} \label{F0xi}
F(\phi)=F_0-\xi\phi^2, 
\end{equation}
where $F_0$ and $\xi$ are nonvanishing constants.
In this case, we can obtain a simple solution, which turns out to be nothing but the constant-roll de Sitter solution obtained in \cite{Motohashi:2019tyj}, showing its uniqueness.

The solution~\eqref{Asol} reads
\begin{equation} \label{Asol-nonmin}
A(\phi) = \frac{2\xi}{4\xi-1}H_0\phi + C H_0^{-\frac{1}{2\xi}+3} \phi^{\frac{1}{2\xi}-1} , 
\end{equation}
where $C$ is a dimensionless integration constant. 
By plugging \eqref{Asol-nonmin} into \eqref{Ueq}, we obtain the potential $U(\phi)$ as
\begin{eqnarray} \label{Usol-nonmin-C}
U(\phi) &=& 3F_0H_0^2 - \frac{\xi (6\xi-1)(16\xi-3)}{(4\xi-1)^2}H_0^2\phi^2 \notag\\
&&-\frac{1}{2}C^2H_0^4\left(\frac{\phi}{H_0}\right)^{\frac{1}{\xi}-2} -\frac{4C\xi(6\xi-1)}{4\xi-1}H_0^4\left(\frac{\phi}{H_0}\right)^{\frac{1}{2\xi}}. 
\end{eqnarray}

Requiring the boundedness of the potential $U(\phi)$ prohibits solution with all $C\ne 0$.
Namely, we can show that if $C\ne 0$, the potential \eqref{Usol-nonmin-C} is unbounded from below for $\xi\ne 0$.
The key is that the coefficient of the third term in \eqref{Usol-nonmin-C} is always negative.
First, when $\xi<0$ or $\xi>\frac{1}{2}$, we have $U\to-\infty$ as $\phi \to 0$ due to the third term.
Second, when $0<\xi\leq \frac{1}{4}$, the third term has the largest positive exponent, and hence $U\to-\infty$ as $\phi \to \infty$.
Finally, when $\frac{1}{4}\leq \xi \leq \frac{1}{2}$, the second term has the largest positive exponent and negative coefficient, and hence $U\to-\infty$ as $\phi \to \infty$.

For $C=0$, the potential~\eqref{Usol-nonmin-C} is given by  
\begin{equation}
U(\phi) = 3F_0H_0^2 - \frac{\xi (6\xi-1)(16\xi-3)}{(4\xi-1)^2}H_0^2\phi^2, \label{Usol-nonmin} \\
\end{equation}
and integrating \eqref{Asol-nonmin}, the inflaton field $\phi(t)$ as a function of time is given by
\begin{eqnarray}
\phi(t) &\propto& H_0 e^{\frac{2\xi}{4\xi-1}H_0t}. \label{phi-nonmin}
\end{eqnarray}
The potential \eqref{Usol-nonmin} is bounded from below for $\xi\leq 0$ and for $\frac{1}{6} \leq \xi \leq \frac{3}{16}$. 
In both cases, this exact de Sitter solution with a massive non-minimally coupled scalar field is a late-time attractor.

Let us remark on the relation to the previous works.
The solution with $\xi<0$ was known long ago \cite{Barrow:1990nv,Sahni:1998at}, and a special case of the conformal coupling $\xi=\frac{1}{6}$ and constant potential was studied in \cite{Kofman:2007tr}.
The solutions of all cases fall into the constant-roll de Sitter solution derived in \cite{Motohashi:2019tyj}.
In \cite{Motohashi:2019tyj}, we started from the constant-roll condition~\eqref{cr-con} and found the de Sitter solution with the constant-roll parameter $\beta = \frac{2\xi}{4\xi-1}$ for a wider class of theory with $F(\phi)=\frac{\beta}{2(1-2\beta)} \phi^2 + c_1 \beta \phi^{\frac{1}{\beta}} + c_2$.
On the other hand, in the above derivation, we start from the potential~\eqref{Ueq} with \eqref{Asol} that accommodates all the de Sitter solutions, and obtain the constant-roll solution.
This shows that the constant-roll de Sitter solution found in \cite{Motohashi:2019tyj} is the unique exact de Sitter solution in the case of a massive non-minimally coupled scalar field with \eqref{F0xi} and bounded potential.

\bibliographystyle{JHEPmod}
\bibliography{refs}
\end{document}